\documentclass{article}

\usepackage{PRIMEarxiv}
\usepackage{hyperref}
\usepackage{graphicx}
\usepackage{listings}
\usepackage{natbib}
\usepackage{url}
\usepackage[export]{adjustbox}

\hypersetup{hidelinks}

\usepackage[dvipsnames]{xcolor}
\definecolor{codeblack}{rgb}{0,0,0}
\definecolor{codegreen}{rgb}{0,0.6,0}
\definecolor{codegray}{rgb}{0.5,0.5,0.5}
\definecolor{codepurple}{rgb}{0.58,0,0.82}
\definecolor{backcolour}{rgb}{0.95,0.95,0.92}

\lstdefinestyle{mystyle}{
    backgroundcolor=\color{backcolour},   
    commentstyle=\color{codegreen},
    keywordstyle=\color{codeblack},
    numberstyle=\tiny\color{codegray},
    stringstyle=\color{codepurple},
    basicstyle=\ttfamily\footnotesize\bfseries,
    breakatwhitespace=false,         
    breaklines=true,                 
    captionpos=t,                    
    keepspaces=true,      
    showspaces=false,                
    showstringspaces=false,
    showtabs=false,                  
    tabsize=2
}
\title{\href{https://github.com/mickcrosse/PERMUTOOLS}{PERMUTOOLS: A MATLAB Package for\\Multivariate Permutation Testing}}

\author{
Michael J. Crosse$^{1,2,3*}$ \quad John J. Foxe$^{3,4}$ \quad Sophie Molholm$^{3,4}$ \\
$^1$Segotia, Galway, Ireland \quad $^2$Trinity College Dublin, Ireland \\
$^3$Albert Einstein College of Medicine, NY \quad $^4$University of Rochester, NY \\
$^{*}$\texttt{crossemj@tcd.ie}
}

\begin{document}

\maketitle

\begin{abstract}
Statistical hypothesis testing and effect size measurement are routine parts of quantitative research. Advancements in computer processing power have greatly improved the capability of statistical inference through the availability of resampling methods. However, many of the statistical practices used today are based on traditional, parametric methods that rely on assumptions about the underlying population. These assumptions may not always be valid, leading to inaccurate results and misleading interpretations. Permutation testing, on the other hand, generates the sampling distribution empirically by permuting the observed data, providing distribution-free hypothesis testing. Furthermore, this approach lends itself to a powerful method for multiple comparison correction — known as max correction — which is less prone to type II errors than conventional correction methods. Parametric methods have also traditionally been utilized for estimating the confidence interval of various test statistics and effect size measures. However, these too can be estimated empirically using permutation or bootstrapping techniques. Whilst resampling methods are generally considered preferable, many popular programming languages and statistical software packages lack efficient implementations. Here, we introduce PERMUTOOLS, a MATLAB package for multivariate permutation testing and effect size measurement.
\end{abstract}

\section*{Background}
\subsection*{Hypothesis testing}
For over a century, researchers have relied on parametric statistical procedures for conducting hypothesis testing, such as the famous Student's \textit{t}-test \citep{student1908}. However, parametric testing was developed out of the necessity to make inferences about the null distribution, as it was impractical to generate it empirically. Today, it is possible to do so using permutation tests. Permutation tests work by permuting the observed data in an appropriate manner to compute the empirical distribution of the test statistic of interest — known as the permutation distribution — which approaches the null distribution \citep{fisher1935}. From the permutation distribution, we can estimate the confidence interval (CI) of the test statistic by computing the corresponding percentiles (e.g. 2.5\% and 97.5\% percentiles for 95\% CI). We can also estimate the probability of observing such a result by chance (i.e. the \textit{p}-value) by calculating the proportion of the permutation distribution that is greater than or equal to the magnitude of the test statistic (Fig. \ref{fig:pdist}). The more permutations generated, the more accurate the estimate. Typically, permutations in the order of several thousand are required to obtain reliable \textit{p}-values, which is computationally trivial for modern computers \citep{ernst2004}.

Permutation testing can be applied to any statistical test. As no assumptions are made about the shape of the underlying distribution, permutation tests provide distribution-free, nonparametric hypothesis testing without the need for rank-transformations \citep{holt2023}. For independent samples, permutation tests have been shown to be relatively insensitive to differences in population variance when samples of equal size are used \citep{murphy1967, groppe2011b}. Moreover, permutation tests have been shown to be more accurate than parametric tests in fields such as biomedical research \citep{ludbrook1998}, and yield more reliable \textit{p}-values and thus are more likely to produce replicable results \citep{noguchi2021}. Despite the widespread use of significance thresholding, it is recommended to treat \textit{p}-values as a continuous measure \citep{mcshane2019}. However, their continuous value does not indicate the strength of an effect, only its likelihood. PERMUTOOLS offers permutation testing and confidence interval estimation for a range of statistical tests, including the ANOVA (one-way, two-way), \textit{t}-test (one-sample, paired-sample, two-sample), \textit{F}-test (two-sample), \textit{Z}-test (one-sample), and correlation test (Pearson, Spearman, rankit). It does not output dichotomous test results based on an arbitrary significance threshold to discourage this practice. 

\subsection*{Correcting for multiple comparisons}
When conducting multiple hypothesis tests simultaneously, there is an increased risk of false discoveries or type I errors. However, many of the traditional correction methods used to control family-wise error rate (FWER) tend to be overly conservative, resulting in increased type II errors (e.g. Bonferroni correction, Holm–Bonferroni method). Researchers are often faced with the delicate task of controlling the trade-off between type I and type II errors. In the biomedical field, researchers tend to preference controlling for type I errors, because the consequences of false positives can be detrimental, e.g. the introduction of an ineffective new therapy or treatment \citep{ludbrook1998}. This inherent bias, along with the use of overly conservative correction methods, can be limiting in the pursuit of scientific progress.

\begin{figure}
  \includegraphics[scale=0.8]{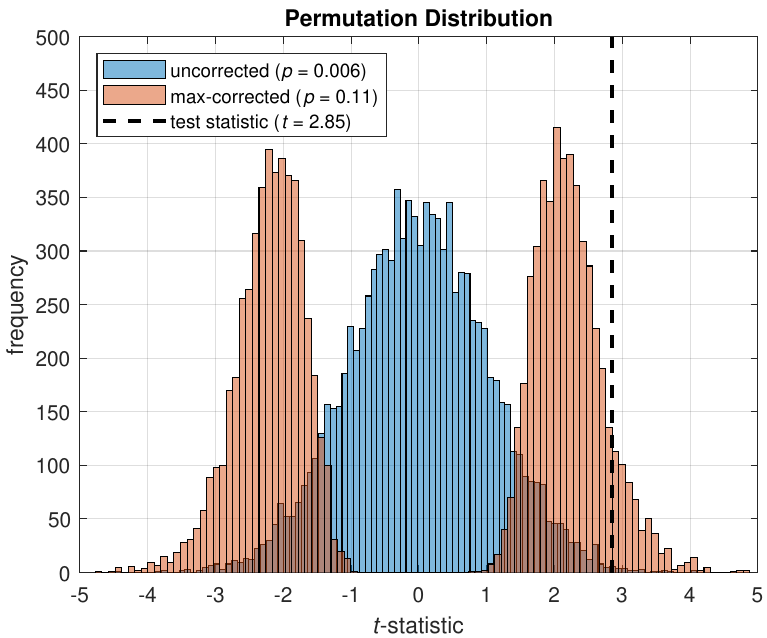}
  \centering
  \caption{Permutation distributions for two-tailed tests based on the \textit{t}-statistic (Blue: uncorrected, Red: max-corrected). Multivariate data were randomly generated to simulate two independent samples, both consisting of 20 variables (i.e. adjusted for 20 comparisons), each with 30 observations (see example code below). The uncorrected permutation distribution shown is that generated for the third test (i.e. variable 3).}
  \label{fig:pdist}
\end{figure}

Another advantage of permutation tests is that they can utilize a powerful technique for correcting for multiple comparisons — known as max correction — which is less prone to type II errors than conventional correction methods \citep{blair1993, westfall1993}. Max correction (also known as \textit{t}\textsubscript{max} correction \citep{blair1994} or joint correction \citep{boca2014}) works as follows: on each permutation of the data, a separate test statistic is computed for each variable and the maximum absolute value (or most extreme positive or negative value) is taken across all variables. Repeating this procedure thousands of times produces a single permutation distribution against which the actual test statistic is compared (Fig. \ref{fig:pdist}). Thus, taking the maximum across more variables naturally produces a more conservative permutation distribution. This highly intuitive approach provides strong control of FWER, even for small sample sizes \citep{gondan2010, groppe2011a, rousselet2023}. Max correction has been used in various scientific disciplines, including the study of electrophysiological data \citep{blair1993, groppe2011a, groppe2011b} and human behavioural data \citep{gondan2010, shaw2020, crosse2022}. PERMUTOOLS automatically applies max correction to multivariate data, unless specified otherwise.

\subsection*{Measuring effect size}
Effect size measurement is an equally important part of inferential statistics \citep{hentschke2011}. Today, most scientific journals require that authors report the size of an effect, and not just its dichotomous existence. A common effect size measure used in research is the standardised mean difference, known as Cohen's \textit{d} \citep{cohen1969}. Standardised effect sizes have the advantage of being metric-free, meaning that they can be directly compared across different studies. For independent samples, Cohen's \textit{d} uses the pooled standard deviation when they are assumed to have equal variances, and an unpooled estimate when this cannot be assumed. For independent samples with significantly different variances, an estimate based on the control sample's variance can be used, known as Glass' delta \citep{glass1976}. In the case of ordinal data, an alternative formulation known as Cliff's \textit{d} should be used \citep{cliff1993}. PERMUTOOLS gives the option to implement any of the above standardised effect size measures, as well as several unstandardised measures.

It is also typical to report the confidence interval of an effect size. Whilst they have traditionally been calculated using parametric methods, these too can be estimated empirically by generating the sampling distribution using a resampling procedure known as bootstrapping. Bootstrapping is fundamentally different to permutation testing in that it resamples with replacement, whereas permutation testing resamples without replacement. PERMUTOOLS uses an efficient bootstrapping algorithm that is optimised for multivariate data analysis.

\subsection*{Correcting for sample size}
As with all statistical tests, the larger the sample size, the more accurately it can describe the population. However, in certain fields such as cognitive science, researchers are often required to work with relatively small ($n<50$) sample sizes due to the limited availability of test subjects or other logistical study constraints. This can hinder the researcher's ability to measure the true, unbiased size of an effect. Despite the advantages of standardised effect size measures, metrics such as Cohen's \textit{d} have been shown to have an upwards bias of up to about 4\% for sample sizes of less than 50. This bias is somewhat reduced by using the pooled weighted standard deviation of the samples \citep{hedges1981}. Additionally, a bias correction factor can be applied to the effect size measure and CI, which is approximately equal to $1-3/(4n-9)$ \citep{hedges1985}. When this correction factor is applied, it is usual to refer to the effect size as Hedges' \textit{g}. PERMUTOOLS automatically applies bias correction when calculating Cohen's \textit{d} and Glass' delta (and their CIs), unless specified otherwise.

\section*{Statement of need}
Whilst resampling methods have gained widespread acceptance, many popular programming languages, including MATLAB, lack efficient implementations or have not yet fully integrated them into their core statistical packages. This hinders the adoption of these robust and versatile techniques by researchers, therefore limiting the quality and reliability of quantitative research. To address this need, PERMUTOOLS provides a comprehensive set of functions in the MATLAB programming language for conducting resampling-based inferential statistics that are easy to use and computationally efficient. Moreover, it is optimised for dealing with large, multivariate datasets and offers powerful methods for correcting for multiple comparisons and sample size, making it an invaluable tool for researchers across various fields of quantitative research. PERMUTOOLS offers a range of new features that distinguish it from existing statistical software packages. Some of the key features are described below.

\subsection*{Key features}
\begin{itemize}
  \item Optimised Resampling Algorithms: PERMUTOOLS utilizes efficient implementations of resampling algorithms that are optimised for multivariate data, ensuring efficient processing of even large datasets with negligible compute times.
  \item Multiple Comparison Correction: PERMUTOOLS implements the powerful max correction method to adjust \textit{p}-values and CIs for multiple comparisons when dealing with multivariate data, reducing the risk of both type I and type II errors.
  \item Sample Size Correction: PERMUTOOLS applies a bias correction factor to the relevant standardised effect size measures and their CIs to adjust for any inherent inflation due to sample size, ensuring less biased estimates for small ($n<50$) samples.
  \item Multivariate Processing: PERMUTOOLS handles multivariate datasets with ease, allowing for multiple tests to be performed simultaneously, as well as the option to perform pairwise comparisons between every combination of variables in a matrix (e.g. correlation matrix).
  \item Continuous Framework: PERMUTOOLS does not output test results under the traditional dichotomous decision-based framework (i.e. $H=0,1$). Instead, it outputs various quantitative measures, encouraging researchers to interpret their results in a continuous and holistic manner.
\end{itemize}

\section*{Using PERMUTOOLS}
The PERMUTOOLS functions are designed to mimic the API of the equivalent parametric functions in MATLAB, with the addition of the prefix `permu' at the beginning of the function name. For example, to conduct a permutation test based on the \textit{t}-statistic, the usual function \verb|ttest()| becomes \verb|permuttest()| etc. The input and output arguments are the same as before with the exception that the first output variable is always the test statistic (and not a dichotomous test result). An additional output variable containing the sampling distribution is included, as well as additional resampling-related input arguments that are described in the help documentation of each function. Unlike many existing statistical software packages, PERMUTOOLS uses a consistent input/output argument framework across all its functions to optimise usability.

\subsection*{Example}
The following example illustrates typical usage of the PERMUTOOLS toolbox. Here, we compare the means of two independent multivariate samples using permutation tests based on the \textit{t}-statistic with max correction, and contrast the results with the equivalent parametric tests in MATLAB (i.e. two-sample \textit{t}-tests). We then measure the associated effect sizes based on the bias-corrected standardised mean difference (i.e. Hedges' \textit{g}), as well as the corresponding 95\% CIs using both a bootstrapping and parametric approach.

First, we generate random multivariate data for two independent samples X and Y. Both samples have 20 variables, each with a mean value of approximately $0$, except for the first 10 variables of Y which have a mean value of approximately $-1$. Each variable contains 30 observations.

\begin{lstlisting}[language=Octave]

  % Generate random data
  rng(42);
  x = randn(30,20);
  y = randn(30,20);
  y(:,1:10) = y(:,1:10)-1;
  \end{lstlisting}

Because we generated the random multivariate data from the same (normal) distribution, we can assume that their variances are approximately equal. If we could not assume this, we would first conduct a two-tailed test of variance based on the \textit{F}-statistic using PERMUTOOLS' \verb|permuvartest2()| function. Thus, we can proceed using the standard Student's \textit{t}-statistic (as opposed to Welch's \textit{t}-statistic). The two-sample permutation tests are implemented using PERMUTOOLS' \verb|permuttest2()| function and the equivalent parametric tests are implemented using MATLAB's \verb|ttest2()| function. By default, max correction is applied to the permutation tests.

\begin{lstlisting}[language=Octave]

  % Run MATLAB's two-sample parametric t-test
  [h1,p1,ci1,stats1] = ttest2(x,y);

  % Run PERMUTOOLS' two-sample permutation t-test
  [t2,p2,ci2,stats2] = permuttest2(x,y);
  \end{lstlisting}

To measure the corresponding effect sizes, we compute Hedges' \textit{g} and the 95\% CIs using an efficient bootstrapping procedure. Effect size measures and bootstrapped CIs are calculated using PERMUTOOLS' \verb|booteffectsize()| function. The equivalent parametric measures are calculated using MATLAB's \verb|meanEffectSize()| function. By default, bias-correction is applied in both cases.

\begin{lstlisting}[language=Octave]

  % Run MATLAB's parametric effect size analysis
  d3 = zeros(1,20); 
  ci3 = zeros(2,20);
  <@\textcolor{blue}{for}@> j = 1:20
      stats3 = meanEffectSize(x(:,j),y(:,j),'effect','cohen','paired',0);
      d3(j) = stats3.Effect;
      ci3(:,j) = stats3.ConfidenceIntervals';
  <@\textcolor{blue}{end}@>

  % Run PERMUTOOLS' bootstrapped effect size analysis
  [d4,ci4,stats4] = booteffectsize(x,y,'effect','cohen','paired',0);
  \end{lstlisting}

\begin{figure}
  \adjincludegraphics[scale=0.9, trim={{0.09\width} 0 {0.09\width} 0},clip]{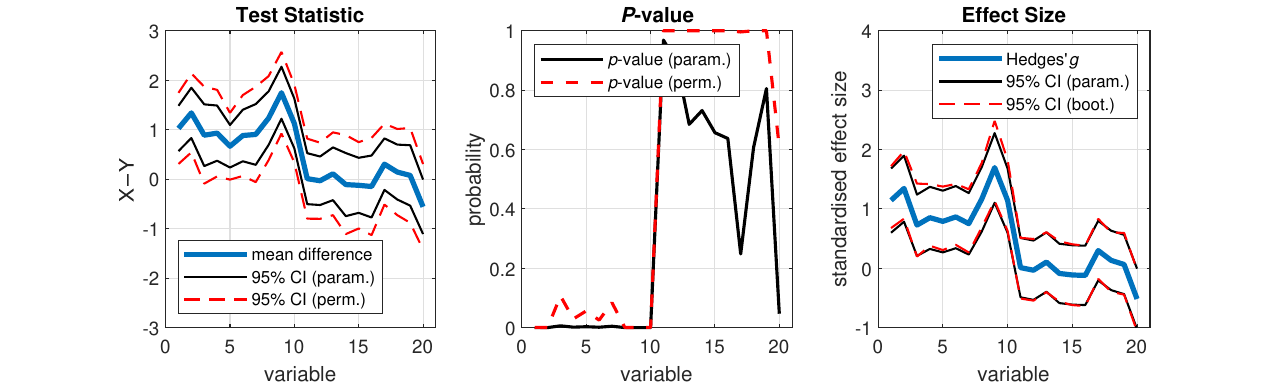}
  \centering
  \caption{Results of the permutation tests and effect size analysis. Left: Test statistic (unstandardised) and 95\% CI (parametric and permutation) for each test. Middle: \textit{P}-value (parametric and permutation) for each test. Right: Effect size (Hedges' \textit{g}) and 95\% CI (parametric and bootstrapped) for each test.}
  \label{fig:example}
\end{figure}

\newpage
To compare the results of our parametric and resampling analyses, we next plot some of the statistics as a function of variable (Fig. \ref{fig:example}). The effect of max correction is clearly visible on the test statistic CIs and the resulting \textit{p}-values, which are consistently more conservative than those of the uncorrected parametric tests (Fig. \ref{fig:example}, left, middle). Importantly, spurious effects observed in the parametric case (e.g. variable 20) did not survive max correction in the permutation tests. In the effect size analysis, the 95\% CIs estimated via bootstrapping appear to approximate the parametric CIs reasonably well (Fig. \ref{fig:example}, right).

\begin{lstlisting}[language=Octave]

  % Set up figure
  figure('Name','Permutation Tests & Effect Size Analysis')
  set(gcf,'color','w')
  xaxis = 1:20; 

  % Plot test statistic & CI
  subplot(1,3,1), hold <@\textcolor{codepurple}{on}@>
  plot(xaxis,stats2.mu,'LineWidth',2.5)
  plot(xaxis,ci1,'k',xaxis,ci2,'--r','LineWidth',1)
  xlim([0,21]), ylim([-3,3]), box <@\textcolor{codepurple}{on}@>, grid <@\textcolor{codepurple}{on}@>
  title('Test Statistic'), xlabel('variable'), ylabel('X-Y')
  legend('mean difference','95% CI (param.)','','95% CI (perm.)')

  % Plot p-value
  subplot(1,3,2), hold <@\textcolor{codepurple}{on}@>
  plot(xaxis,p1,'k',xaxis,p2,'--r','LineWidth',1.5)
  ylim([0,1]), xlim([0,21]), box <@\textcolor{codepurple}{on}@>, grid <@\textcolor{codepurple}{on}@>
  title('{\itP}-value'), xlabel('variable'), ylabel('probability')
  legend('{\itp}-value (param.)','{\itp}-value (perm.)') 

  % Plot effect size & CI
  subplot(1,3,3), hold <@\textcolor{codepurple}{on}@>
  plot(xaxis,d4,'LineWidth',2.5)
  plot(xaxis,ci3,'k',xaxis,ci4,'--r','LineWidth',1)
  ylim([-2,6]), xlim([0,21]), box <@\textcolor{codepurple}{on}@>, grid <@\textcolor{codepurple}{on}@>
  title('Effect Size'), xlabel('variable'), ylabel('Hedges'' {\itg}')
  legend('effect size','95% CI (param.)','','95% CI (boot.)')
  \end{lstlisting}

From the above statistical analysis, we can examine the results of each individual pairwise comparison between X and Y, which are contained in the variables output by the PERMUTOOLS functions. Taking the first variables of X and Y as an example, we see that the mean of X\textsubscript{1} ($M = -0.06$, $SD = 0.91$) was significantly greater than the mean of Y\textsubscript{1} ($M = -1.09$, $SD = 0.86$), even after adjusting for multiple comparisons ($t(58) = 4.49$, $p = 0.0008$, Hedges' $g = 1.14$, 95CI $[0.68, 1.72]$). We recommend always reporting the absolute values of the tests as shown here, in particular, with the effect size and CI included.

\section*{Future work}
PERMUTOOLS is under active development, and new functions and features are constantly being added to it as needed. Future work aims to expand the toolbox to provide new permutation-based tests, including repeated measures ANOVAs (one-way, two-way), and multi-way ANOVAs (\textit{n}-way), as well as expanding the functionality of existing ANOVA tests to be able to deal with unbalanced and multivariate samples. Whilst permutation tests have been shown to outperform nonparametric tests based on rank transformations \citep{holt2023}, there are certain situations where a rank-transformation approach is desirable; for example, when dealing with ordinal data and outliers. To accommodate a wider range of data types, future work aims to develop permutation-based solutions for nonparametric tests based on rank transformations, e.g. Sign test, Wilcoxon signed rank test, Mann-Whitney \textit{U}-test, Kruskal-Wallis test, Friedman test, etc.

PERMUTOOLS is maintained by the corresponding author but accepts contributions from the research community at large. If you would like to contribute to PERMUTOOLS, or request that a specific statistical test or feature be added to it, please email the corresponding author at the email address provided above, or use the formal channels on GitHub, such as creating a pull request or opening an issue.

\section*{Availability}
PERMUTOOLS is freely available to download from GitHub (\url{https://github.com/mickcrosse/PERMUTOOLS}) and is compatible with Windows, Linux, and macOS. Example code demonstrating its usage is included in the examples folder and on the GitHub page.

\section*{Acknowledgments}
We would like to thank David M. Groppe for his helpful correspondences during the development of the toolbox, as well as his open-source code which served as a valuable benchmark. This work was supported by the National Institute of Mental Health of the National Institutes of Health (NIH) under award number R01MH085322 (S.M. and J.J.F.), and the Eunice Kennedy Shriver National Institute of Child Health and Human Development of the NIH under the award numbers P50HD105352, previously U54HD090260 (Rose F. Kennedy Intellectual and Developmental Disabilities Research Center), and P50HD103536 (University of Rochester Intellectual and Developmental Disabilities Research Center).

\section*{Author contributions}
M.J.C. conceived of the toolbox and manuscript, developed the toolbox and example code, and wrote the first draft of the manuscript. S.M. and J.J.F. supervised and edited the manuscript.

\bibliographystyle{plainnat}
\bibliography{references}

\begin{thebibliography}{24}
\providecommand{\natexlab}[1]{#1}
\providecommand{\url}[1]{\texttt{#1}}
\expandafter\ifx\csname urlstyle\endcsname\relax
  \providecommand{\doi}[1]{doi: #1}\else
  \providecommand{\doi}{doi: \begingroup \urlstyle{rm}\Url}\fi

\bibitem[Blair and Karniski(1993)]{blair1993}
R~Clifford Blair and Walt Karniski.
\newblock An alternative method for significance testing of waveform difference potentials.
\newblock \emph{Psychophysiology}, 30\penalty0 (5):\penalty0 518--524, 1993.

\bibitem[Blair et~al.(1994)Blair, Higgins, Karniski, and Kromrey]{blair1994}
R~Clifford Blair, James~J Higgins, Walt Karniski, and Jeffrey~D Kromrey.
\newblock A study of multivariate permutation tests which may replace hotelling's t2 test in prescribed circumstances.
\newblock \emph{Multivariate Behavioral Research}, 29\penalty0 (2):\penalty0 141--163, 1994.
\newblock ISSN 0027-3171.

\bibitem[Boca et~al.(2014)Boca, Sinha, Cross, Moore, and Sampson]{boca2014}
Simina~M Boca, Rashmi Sinha, Amanda~J Cross, Steven~C Moore, and Joshua~N Sampson.
\newblock Testing multiple biological mediators simultaneously.
\newblock \emph{Bioinformatics}, 30\penalty0 (2):\penalty0 214--220, 2014.

\bibitem[Cliff(1993)]{cliff1993}
Norman Cliff.
\newblock Dominance statistics: Ordinal analyses to answer ordinal questions.
\newblock \emph{Psychological bulletin}, 114\penalty0 (3):\penalty0 494, 1993.

\bibitem[Cohen(1969)]{cohen1969}
Jacob Cohen.
\newblock \emph{Statistical power analysis for the behavioral sciences}.
\newblock Academic press, 1969.

\bibitem[Crosse et~al.(2022)Crosse, Foxe, Tarrit, Freedman, and Molholm]{crosse2022}
Michael~J Crosse, John~J Foxe, Katy Tarrit, Edward~G Freedman, and Sophie Molholm.
\newblock Resolution of impaired multisensory processing in autism and the cost of switching sensory modality.
\newblock \emph{Communications biology}, 5\penalty0 (1):\penalty0 601, 2022.

\bibitem[Ernst(2004)]{ernst2004}
Michael~D Ernst.
\newblock Permutation methods: a basis for exact inference.
\newblock \emph{Statistical Science}, pages 676--685, 2004.

\bibitem[Fisher(1935)]{fisher1935}
Ronald~A Fisher.
\newblock \emph{The Design of Experiments}.
\newblock Hafner, New York, 1935.

\bibitem[Glass(1976)]{glass1976}
Gene~V Glass.
\newblock Primary, secondary, and meta-analysis of research.
\newblock \emph{Educational researcher}, 5\penalty0 (10):\penalty0 3--8, 1976.

\bibitem[Gondan(2010)]{gondan2010}
Matthias Gondan.
\newblock A permutation test for the race model inequality.
\newblock \emph{Behavior Research Methods}, 42\penalty0 (1):\penalty0 23--28, 2010.
\newblock ISSN 1554-351X.

\bibitem[Groppe et~al.(2011{\natexlab{a}})Groppe, Urbach, and Kutas]{groppe2011a}
David~M Groppe, Thomas~P Urbach, and Marta Kutas.
\newblock Mass univariate analysis of event‐related brain potentials/fields i: A critical tutorial review.
\newblock \emph{Psychophysiology}, 48\penalty0 (12):\penalty0 1711--1725, 2011{\natexlab{a}}.
\newblock ISSN 0048-5772.

\bibitem[Groppe et~al.(2011{\natexlab{b}})Groppe, Urbach, and Kutas]{groppe2011b}
David~M Groppe, Thomas~P Urbach, and Marta Kutas.
\newblock Mass univariate analysis of event-related brain potentials/fields ii: Simulation studies.
\newblock \emph{Psychophysiology}, 48\penalty0 (12):\penalty0 1726--1737, 2011{\natexlab{b}}.

\bibitem[Hedges(1981)]{hedges1981}
Larry~V Hedges.
\newblock Distribution theory for glass's estimator of effect size and related estimators.
\newblock \emph{journal of Educational Statistics}, 6\penalty0 (2):\penalty0 107--128, 1981.

\bibitem[Hedges and Olkin(1985)]{hedges1985}
Larry~V Hedges and Ingram Olkin.
\newblock \emph{Statistical methods for meta-analysis}.
\newblock Academic press, 1985.

\bibitem[Hentschke and St{\"u}ttgen(2011)]{hentschke2011}
Harald Hentschke and Maik~C St{\"u}ttgen.
\newblock Computation of measures of effect size for neuroscience data sets.
\newblock \emph{European Journal of Neuroscience}, 34\penalty0 (12):\penalty0 1887--1894, 2011.

\bibitem[Holt and Sullivan(2023)]{holt2023}
Charles~A Holt and Sean~P Sullivan.
\newblock Permutation tests for experimental data.
\newblock \emph{Experimental Economics}, pages 1--38, 2023.

\bibitem[Ludbrook and Dudley(1998)]{ludbrook1998}
John Ludbrook and Hugh Dudley.
\newblock Why permutation tests are superior to t and f tests in biomedical research.
\newblock \emph{The American Statistician}, 52\penalty0 (2):\penalty0 127--132, 1998.

\bibitem[McShane et~al.(2019)McShane, Gal, Gelman, Robert, and Tackett]{mcshane2019}
Blakeley~B McShane, David Gal, Andrew Gelman, Christian Robert, and Jennifer~L Tackett.
\newblock Abandon statistical significance.
\newblock \emph{The American Statistician}, 73\penalty0 (sup1):\penalty0 235--245, 2019.

\bibitem[Murphy(1967)]{murphy1967}
Brian~P Murphy.
\newblock Some two-sample tests when the variances are unequal: a simulation study.
\newblock \emph{Biometrika}, 54\penalty0 (3-4):\penalty0 679--683, 1967.

\bibitem[Noguchi et~al.(2021)Noguchi, Konietschke, Marmolejo-Ramos, and Pauly]{noguchi2021}
Kimihiro Noguchi, Frank Konietschke, Fernando Marmolejo-Ramos, and Markus Pauly.
\newblock Permutation tests are robust and powerful at 0.5\% and 5\% significance levels.
\newblock \emph{Behavior Research Methods}, 53\penalty0 (6):\penalty0 2712--2724, 2021.

\bibitem[Rousselet(2023)]{rousselet2023}
Guillaume~A Rousselet.
\newblock Using cluster-based permutation tests to estimate meg/eeg onsets: how bad is it?
\newblock \emph{bioRxiv}, pages 2023--11, 2023.

\bibitem[Shaw et~al.(2020)Shaw, Freedman, Crosse, Nicholas, Chen, Braiman, Molholm, and Foxe]{shaw2020}
Luke~H Shaw, Edward~G Freedman, Michael~J Crosse, Eric Nicholas, Allen~M Chen, Matthew~S Braiman, Sophie Molholm, and John~J Foxe.
\newblock Operating in a multisensory context: Assessing the interplay between multisensory reaction time facilitation and inter-sensory task-switching effects.
\newblock \emph{Neuroscience}, 436:\penalty0 122--135, 2020.

\bibitem[Student(1908)]{student1908}
Student.
\newblock The probable error of a mean.
\newblock \emph{Biometrika}, 6\penalty0 (1):\penalty0 1--25, 1908.

\bibitem[Westfall and Young(1993)]{westfall1993}
Peter~H Westfall and S~Stanley Young.
\newblock \emph{Resampling-based multiple testing: Examples and methods for p-value adjustment}, volume 279.
\newblock John Wiley \& Sons, 1993.

\end{thebibliography}

\end{document}